# Coupling Between Local and Global Oscillations in Palladium-Catalysed Methane Oxidation

Yuxiong Hu[1#], Jianyu Hu[2#], Mengzhao Sun[1#], Aowen Li[3], Shucheng Shi[1], P. J. Hu[1,4], Wu Zhou[3], Marc-Georg Willinger[5], Dan Zhou[6,7]*, Zhi Liu[1,8], Xi Liu[9], Wei-Xue Li[2,10,11]*, Zhu-Jun Wang[1,5,8]*

[1]School of Physical Science and Technology, ShanghaiTech University, Shanghai, China.
[2]Hefei National Research Center for Physical Sciences at the Microscale, University of Science and Technology of China, Hefei, China.
[3]School of Physical Sciences and CAS Key Laboratory of Vacuum Physics, University of Chinese Academy of Sciences, Beijing, People's Republic of China.
[4]School of Chemistry and Chemical Engineering, Queen's University Belfast, Belfast, UK.
[5]Department of Chemistry, Technical University of Munich, Garching, Germany.
[6]Department of Materials Science, Leibniz-Institut für Kristallzüchtung, Berlin, Germany.
[7]DENSsolutions B.V., Delft, The Netherlands.
[8]Center for Transformative Science, Shanghai Tech University, Shanghai, China.
[9]School of Chemistry and Chemical Engineering, In-situ Center for Physical Sciences, Frontiers Science Center for Transformative Molecules, Shanghai Jiao Tong University, Shanghai, China
[10]Key Laboratory of Precision and Intelligent Chemistry, School of Chemistry and Materials Science, Collaborative Innovation Center of Chemistry for Energy Materials (iChEM), University of Science and Technology of China, Hefei, China.
[11]Hefei National Laboratory, University of Science and Technology of China, Hefei, Anhui, China.

**Abstract**

The interplay between order and disorder is crucial across various fields, especially in understanding oscillatory phenomena. Periodic oscillations are frequently observed in heterogeneous catalysis, yet their underlying mechanisms need deeper exploration. Here, we investigate how periodic oscillations arise during methane oxidation catalysed by palladium nanoparticles (Pd NPs), utilizing a suite of complementary operando techniques across various spatial scales. We found that reaction intensity and collective dynamic modes can be tuned by the reactant gas-flow rate. At lower gas-flow rates, we observed periodic facet reconstruction of Pd NPs correlated with repeated bubbling behaviour at the Pd/PdO interface, without evident global oscillatory responses. Conversely, at higher gas-flow rates, Pd NPs undergo chaotic transformations between metallic and oxidized states, resulting in overall oscillation. Integrating our observations at different gas-flow rates, we attributed the emergence of global oscillation to thermal coupling regulated by gas flow and connected local and global dynamics through a weak synchronization mechanism. This work demonstrates the correlations between open surfaces and interfaces, chaos and regularity, and dissipative processes and coupling behaviour. Our findings offer critical insights into the complexity behind catalytic oscillations and provide guidance for modulating oscillatory behaviours in catalytic processes, with significant implications for both science and industry.

**Introduction**

Heterogeneous catalytic reactions, when operating away from thermodynamic equilibrium[1], frequently exhibit self-organizing phenomena such as chaos[2,3], spatio-temporal patterns[4-7], and oscillatory behaviours[8-10], known as "dissipative structures" per Prigogine's terminology[11]. These structures arise in coupled systems with significant interplay between different agents, modulated by internal interactions and external exchange of mass and energy[12-14]. This principle is recognized in systems like nanocrystal self-assembly[15,16], cell movement[17,18], oscillator synchronization[19], and magnetic material structure formation[20,21]. However, few studies have applied this principle to understand the dissipative structures in heterogeneous catalytic reactions composed of NPs or clusters, which represent one of the most academically significant and industrially relevant systems[22,23].

Among dissipative structures, self-sustained oscillations are common in metal-catalysed reactions[5,6,8,24-27] and are crucial for developing efficient, never-deactivating catalysts. These phenomena result from nonlinear interactions and feedback mechanisms[28], integrating temporal and spatial factors with mass and heat exchanges across multiple scales[5,6,29]. Previous reports attribute oscillatory behaviours to repeated reaction steps in a catalytic cycle[30,31] or periodic transitions between different surface structures of catalysts[5,32-34]. Despite extensive study, the mechanisms by which NP dynamics foster self-sustained global oscillations and their specific structure-activity relationships remain elusive, primarily due to the lack of temporally and spatially resolved information on catalyst structures under relevant conditions. Moreover, NPs are often treated as identical replicas[35-37], neglecting their subtle interactions with neighbouring particles[33,34,38]. Consequently, the driving forces behind the spontaneous formation of self-sustained global oscillations are yet to be fully understood.

Here, we investigate the mechanisms of oscillation in heterogeneous catalytic systems using $CH_4$ oxidation over Pd NPs. Understanding this exothermic reaction is beneficial for converting

methane—a significant greenhouse gas[39]—into syngas, a crucial building block for fuels and chemicals[40]. We combined multiple *operando* characterization methods to observe chemical oscillations in gas-solid heterogeneous catalytic systems at different spatial levels. Specifically, we employ *operando* transmission electron microscopy (TEM) for real-time[33], real-space imaging of catalyst NPs at a local view, alongside online mass spectrometry (MS) and near-ambient pressure X-ray photoelectron spectroscopy (NAPXPS) to monitor the global atmosphere composition and real-time chemical state of catalysts during the reaction. Integrating these *in situ* characterization methods allows us to track the catalyst's morphological transformations and catalytic reactivity simultaneously. Additionally, diverse quasi *in situ* characterization methods, including focused ion beam-scanning electron microscopy (FIB-SEM), MS, and electron energy loss spectrum (EELS), were applied to investigate the quenching features of catalysts.

At micrometre scales, TEM images and selected area electron diffraction (SAED) patterns provide statistical phase distribution of catalyst particles. Theoretical simulations further elucidate structure-performance relationships of single NPs and the collective dynamics of the entire particle group. The system's dissipative behaviour is controlled by a temperature controller and gas supply system of a microelectromechanical system (MEMS)-based microreactor for *operando* TEM. Gas-flow rates modulate the dynamic intensity among Pd NPs. At low gas-flow rates, periodic surface reconstruction of Pd NPs results from water adsorption/desorption due to interfacial bubbling, keeping the system in an apparent static state. Higher gas-flow rates enhance NPs' thermal interactions, leading to coupling among Pd NPs and the emergence of global oscillation. Pd NPs also demonstrate self-motivated movement, indicating greater catalytic reactivity compared to low gas-flow rates. Real-time tracking of temperature and power revealed the interplay between heat and gas flow, showing real-time global variations accompanied by online MS signals. Mesoscale observations reveal weak synchronization within these heat and flow fields, forming an effective mesoscale oscillator and eventually achieving global oscillation.

Our approach correlates atomic-scale observations with broader multi-signal insights, elucidating the relationship between local NP behaviours and global dynamics in self-sustained oscillation during Pd-catalysed methane oxidation. Our findings guide the use of oscillatory behaviours in catalytic processes, with important implications for science and the chemical industry. Additionally, our work uncovers hidden order behind apparent randomness and chaos in chemical processes, inspiring further research into the complexity of heterogeneous catalytic reactions. The self-driven dynamic mode of NP catalysts is analogous to living cells[41], or molecule motors[42,43], bridging the gap between living matter and abiotic systems and reflecting the universality of dissipative structures in coupled systems.

**Results and Discussion**

**Periodic Reconstruction of the Pd Facets**

At relatively low gas-flow rates (~0.0011 sccm, $P_{total}$ = 155 mbar), we observed periodic reconstruction of the Pd facets, similar to surface dynamics in other heterogeneous catalytic systems[33]. TEM observations, combined with a fast Fourier transform (FFT) pattern, indicate that the reconstruction occurs at the Pd (110) planes (**Figure 1a-1b**). The presence of other crystal facets is further validated from different views (highlighted with coloured arrows in **Figure 1a**).

**Figure 1c-1f** show the localized periodic surface reconstruction of a Pd NP within one cycle (~7 seconds) where green and red arrows represent the direction of surface reconstruction. **Figure 1g** depicts the initial state of the coming cycle. We reconstructed the time-sequence stacks of operando TEM images using a time-parameterized colour bar (**Figure 1h**). The corresponding evolution of TEM contrast signal intensity for this particle along the [1$\bar{1}$0] direction (highlighted by the green rectangle in **Figure 1c-1g**) is plotted in **Figure 1i**. In each cycle, the Pd NP morphology becomes more protruded along the [1$\bar{1}$0] direction, lasting a few seconds, while the return to the initial morphology takes about 0.2 seconds (**Figure 1i**). The ratio of {110} facet at in-plane direction decreases continuously during the former long process, as shown by the Wulff models (the black rectangle region in **Figure 1j-1k**). No detectable global temperature or heating power oscillation is monitored by the MEMS chip during this morphological evolution (**Figure 1l**), indicating that the system remains in apparent thermal equilibrium.

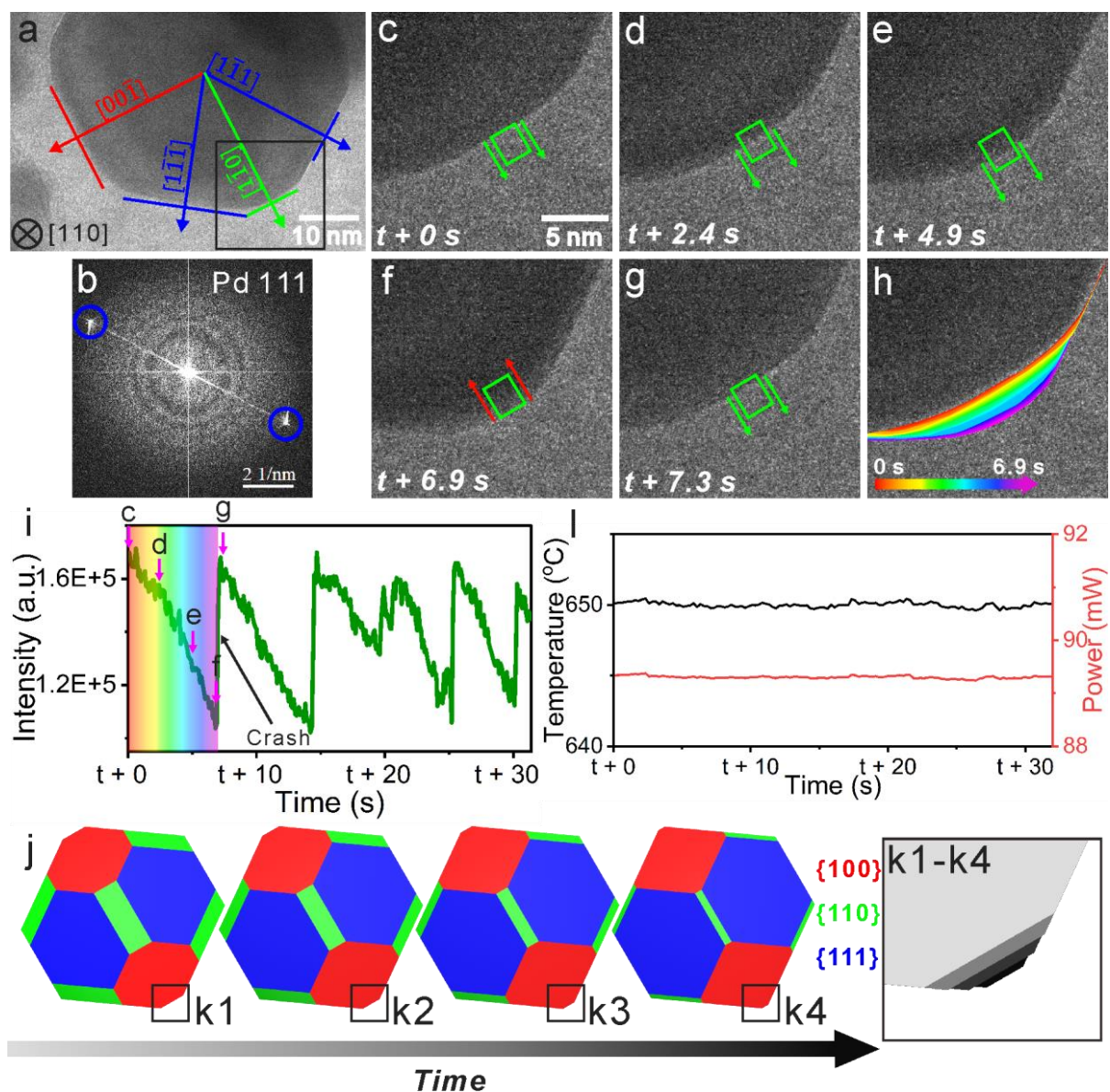

**Figure 1. Periodic Reconstruction Behaviour of Pd NPs Observed via *Operando* TEM**. (a) The Pd NP exhibits Wulff structure morphology at a CH$_4$/O$_2$ ratio of ~2/3 (P$_{total}$ = 155 mbar) and 650°C with a gas-flow rate of ~0.0011 sccm. The out-of-plane direction set as the [110] zone axis where the red, blue, and green arrows are perpendicular to the corresponding-coloured planes of $(00\bar{1})$, $(1\bar{1}\bar{1})$, and $(1\bar{1}0)$, respectively. (b) The FFT pattern of (a) reveals a d-spacing corresponding to the Pd $(1\bar{1}1)$ plane. (c-g) The sequence of images is magnified from the black rectangle in (a), showing the corner growth and collapse process, where each moment is marked in (i) with pink arrows. (h) Shape evolution of the palladium corner during growth, reproduced as a color-coded superposition of outlines abstracted from images (c-f)

recorded over a 6.9-second interval, with a colour bar set from 0 s to 6.9 s. (i) The intensity evolution over time in the region of green rectangle located in images (c-g), with green arrows predicting the growth direction of the surface and a red arrow for the surface back direction. The sudden increase during intensity evolution corresponds to the back of the surface with a black arrow indication. Here, the colourful background corresponds to (h). (j) Wulff model illustrating the decreasing {110} area according to the facet evolution in images (c-f), where {100}, {110} and {111} are marked with red, green and blue, respectively. The colour bar of time corresponds to the increased contrast of (k1-k4). (k1-k4) Close-up views of the black rectangle region in (j), with colour-continued contrast showing the clear decrease of the $(1\bar{1}0)$ plane. (l) Temperature and power during the entire surface reconstruction period where temperature fluctuations are lower than 0.5°C, accompanied by power changes lower than 0.3 mW, and show no periodicity. (300 kV, electron dose rate: 148 e$^-$/ (Å$^2$·s))

Periodic surface reconstruction of the Pd NP is accompanied by TEM contrast fluctuations in the central area of the observed Pd NP. Increasing pressure makes these TEM contrast fluctuations more pronounced (up to 902 mbar, ~0.0036 sccm). FFT patterns confirm that these weak contrast features form, grow, and eventually vanish at the Pd/PdO interface. Additionally, these features expand only towards the metallic Pd side, suggesting that Pd has better malleability than PdO[44,45]. To investigate the structural characteristics of these features, we constructed atomic models of the Pd/PdO interface based on experimentally observed crystal orientations through searching for optimized configurations. These models show that the structure optimized interfaces are defective with lattice mismatches, and that Pd bulk at the interface may undergo significant deformation. Therefore, we hypothesize that the weak contrast features are inner voids caused by interfacial chemical reactions. To confirm our hypothesis, FIB-SEM technology was used to mill the sample layer by layer, revealing the void morphology inside the particle and confirming that the TEM-observed features are interfacial voids (**Figure 2a** and **Figure 2b-2c**).

The periodic bubbling cycle at Pd/PdO interface coincides with the periodic reconstruction of Pd morphology, as captured by *operando* TEM observations (**Figure 2d-2g**). When the voids burst, the Pd NP rapidly reverts from the sharp Wulff shape to a smoother form. To confirm the synchronization between these two oscillatory behaviours, we stacked time-resolved TEM images of a region (light blue rectangle **h** in **Figure 2d**) containing both the interface and open surface (green and red rectangles respectively in **Figure 2h**). The relative contrast intensity changes for facets reconstruction and void formation /collapse are shown by the red and green curves in **Figure 2i**. The alignment of peaks in the red curve with troughs in the green curve (marked by blue vertical lines) indicates that the void collapse and the instantaneous reformation of Pd NP occur simultaneously, sharing the same period.

Further insights into the correlation between surface reconstruction and interfacial void evolution are provided. The inset of **Figure 2i** (purple-framed inset) shows the contrast evolution over one cycle, with moments marked by blue dashed lines corresponding to the spatial images shown in **Figure 2j** to **2m**. As the voids grow, the Pd NP undergoes anisotropic reconstruction with diminishing {110} facets (**Figure 2j1-2m1**), consistent with TEM contrast images stacking projection information. To avoid the possibility of beam-induced artifacts[46], the impact of the electron beam on the catalyst structure was tested. The results show that the electron beam only contributes to the merging of the Pd/PdO interface, which is detrimental to the bubbling phenomenon.

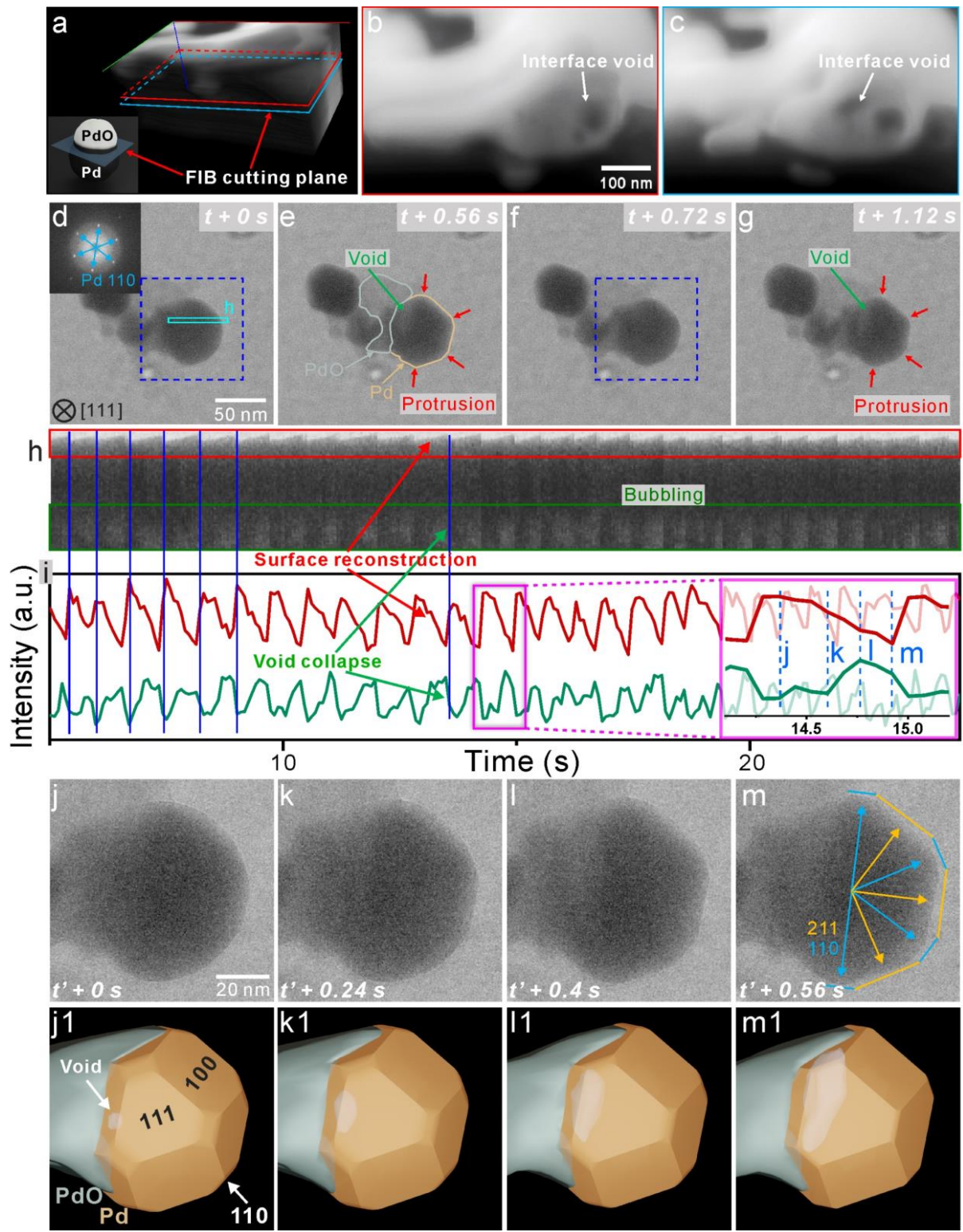

**Figure 2. Bubbling Dynamics of The Pd/PdO Interface.** (a) FIB-SEM tomography of the Pd/PdO interface achieved by layer-by-layer milling. The inset shows the model corresponding

to the FIB-SEM cutting with red arrows indication. (b, c) SEM images corresponding to the red and blue cutting positions in (a), respectively, recording the voids in the Pd/PdO interface during the FIB milling process. White arrows highlight the interfacial void. (d-g) Operando TEM observations of periodic Pd morphology reconstruction at a $CH_4/O_2$ ratio of ~2/3 ($P_{total}$ = 902 mbar) and 650°C with a gas-flow rate of ~0.0036 sccm. The inset in (d) is the FFT, showing the Pd [111] zone axis with Pd <110> directions of azure arrows. Images (e) and (g) show the void region and surface protrusion compared to the blue dashed rectangle region of (d) and (f), with green and red arrows, respectively. The phase division of Pd/PdO is delineated by the lines and arrows of grey and light yellow in (e), respectively, through contrast identification[46-48]. (h) Time-stacking of the light blue rectangle region in (d), showing the sequence of surface reconstruction and the evolution of the inner void in the particle. (i) The TEM contrast time evolution of the Pd NP surface and the Pd/PdO interface region in (h) are indicated by the red and green rectangles, respectively. The red and green arrows indicate the relationship between image features, and corresponding intensities of surface reconstruction and void collapse, respectively. The blue vertical lines show the synchronization between surface back and void collapse. The inset shows a magnified view of the pink rectangle with a semi-transparent background, where each moment from (j-m) is marked with blue dashed lines. (j-m) Particle morphology evolution during one period of oscillation corresponding to the inset in (f). The arrows in (m) indicate the <110> and <211> directions with azure and yellow colours, respectively, where the arrows are vertical to the lines. (j1-m1) 3D models corresponding to (j-m) illustrate void formation and surface reconstruction, with PdO shown in grey and Pd in light yellow. The voids are indicated by white arrows. (80 kV, electron dose rate: 78.9 e$^-$/ (Å$^2$·s))

**Reconstruction Caused by Adsorption and Desorption of H$_2$O Released as Voids Breakup**

To identify the reaction responsible for the interfacial bubbling behaviour, we obtained chemical state information at different depths of the Pd/PdO interface by gradually changing the photon energy in NAPXPS (**Figure 3a-3b**). As the photon energy increases, the H$_2$O signal becomes stronger relative to the surface-adsorbed species OH and O, indicating that the voids contain water[49]. This is further confirmed by the vibrational EELS analysis, which shows a decrease in the OH peak of $H_2O$[50] with increasing distance from the interfacial void (**Figure 3c**). Theoretical simulations are conducted to understand the process of water production and cavity formation at the Pd/PdO interface. This directly reveals that the reaction also occurs at a confined interface, differing from the traditional viewpoint[51,52].

Based on the above observations, we applied first-principle calculation to interpret how the periodically released $H_2O$ from the void influences the morphological changes of Pd NPs. Our simulations, show that the relative surface energy of Pd (110) to Pd (111) facets decreases dramatically when the partial pressure of $H_2O$ increases under a constant oxygen pressure (**Figure 3d**). Since the variation of metal NPs morphology with gas (including $H_2$, $O_2$, $N_2$, etc) adsorption and desorption has been previously reported[53-55], the coupling between interface bubbling and surface reconstruction can be elucidated as follows. During the buildup of voids at the interface, the formed $H_2O$ is trapped. The absorbed $H_2O$ on the open Pd surface is gradually desorbed due to the exothermic reaction of $CH_4$ oxidation, causing the Pd (110) facet with higher surface energy to shrink. Conversely, when voids at the Pd/PdO interface burst, the accumulated $H_2O$ is released and attaches to the Pd surface. This sudden increase in local water pressure causes the Pd nanoparticle to revert from a rounder form to a sharp Wulff structure with a larger (110) facet (**Figure 3e**). Wulff's model of Pd NP, established according to the surface energy of low-index facets (**Figure 3e**), aligns with the observed evolution behaviour of Pd NP (**Figure 1j**). Furthermore, the impact of oxygen adsorption can be excluded, as oxygen adsorption would increase the surface energy ratio of Pd (110) compared to Pd (111) according to simulation results, which is inconsistent with our observations. In summary, the dynamics of open surfaces and closed interfaces are interlinked, collaboratively driving the periodic morphological evolution of Pd NPs. The repeated interfacial bubbling is a local behaviour, while the entire system displays a quasi-static mode without macroscopic oscillation.

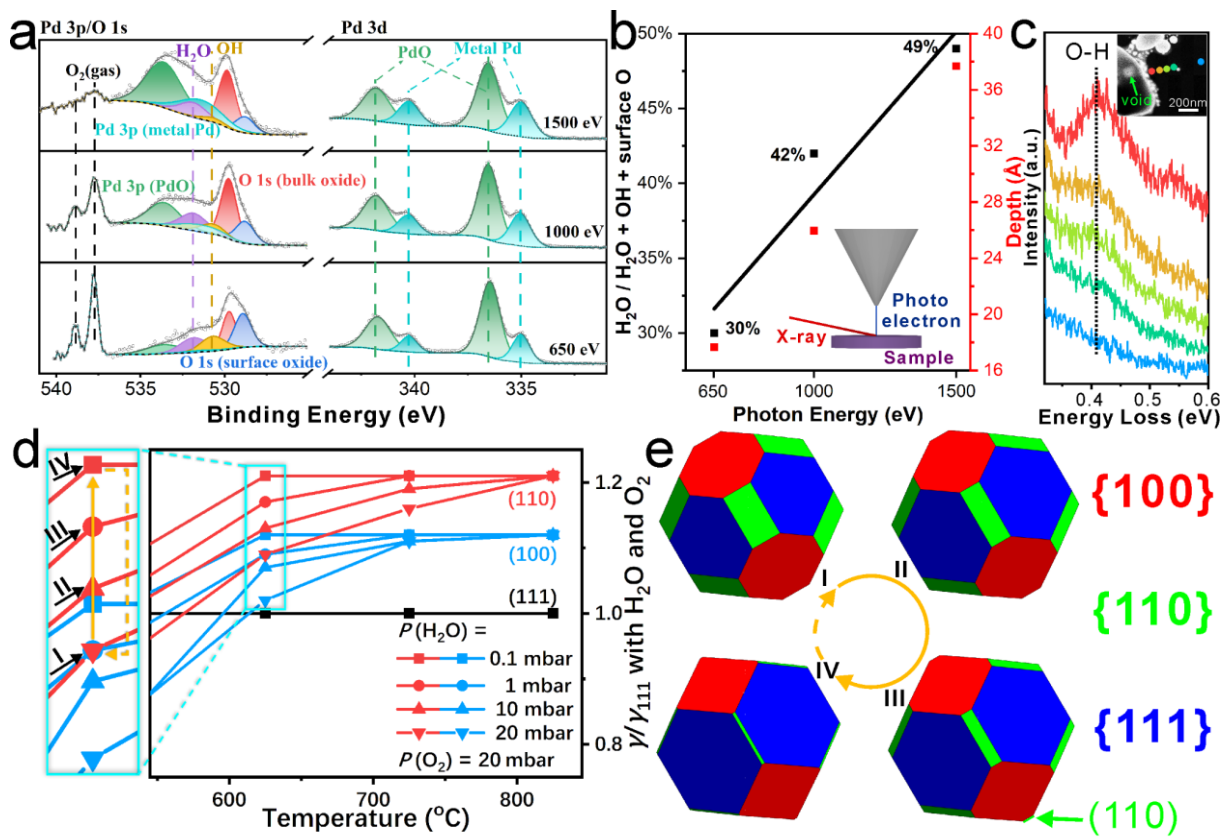

**Figure 3. Confirmation of Content in The Void at The Pd/PdO Interface and Calculations of Surface Energy Variation with Changing Water Pressure**. (a) Pd 3p/O 1s and Pd 3d spectra collected under 0.5 mbar $CH_4$ + 2 mbar $O_2$ at approximately 625°C with photon energies of 650, 1000, and 1500 eV. The Pd 3d spectra suggest the coexistence of metallic Pd and PdO[56], pointed out by green and light blue dashed arrows, respectively. The area ratio of Pd to PdO in the Pd 3p spectrum is consistent with that in the Pd 3d spectrum. Peaks corresponding to $H_2O$ and OH are marked in purple and brown, respectively. Other peaks also correspond to the corresponding colours, respectively. (b) The evolution of the $H_2O$ / ($H_2O$ + OH + surface O) ratio as a function of depth and photon energy. The black line represents the linear fitting of the ratio evolution, and the inset shows a schematic of the experimental apparatus demo. (c) The EELS vibration spectrum, where the 412 meV peak corresponds to the O-H signal from water[50]. The black dashed line indicates the decreasing O-H signal with increasing distance from the void. The insert shows the HAADF image of Pd NPs after the operando TEM experiment, with

different coloured dots indicating EELS collection positions, where the green arrow highlights the position of the void. (d) Diagram of the surface energy ratio for (100) / (111), (110) / (111), and (111) / (111) at $P_{O2}$ = 20 mbar and varying $P_{H2O}$ at different temperatures. The magnified results at 625°C with azure rectangle are used to construct the Wulff model, where the yellow arrow indicates the direction of continual evolution and the yellow dashed arrow displays the quick transformation, forming a loop. The black arrows with Roman numerals point to four stages of red symbol, denoting the variation of (110) / (111) comparison. (e) Wulff models for $P_{H2O}$ = 0.1, 1, 10, and 20 mbar, respectively numbered with Roman numerals in a clockwise sequence highlighted by the yellow curved arrow, at 625°C, as shown in (d). The (110) plane of the in-plane direction is indicated by the green arrow and line in III. As water pressure decreases, the {110} ratio decreases, which is consistent with the observations in **Figure 1**.

**Global Oscillation Induced by High Gas-Flow Rate**

After slightly increasing gas-flow rate (from ~0.0036 sccm to ~0.014 sccm, $P_{total}$ = 902 mbar), the conversion of NPs between the metallic and oxidized phases accelerates significantly, and the Pd/PdO interface begins to migrate. As shown in **Figure 4a-4d**, the Pd/PdO interface moves towards the metallic Pd side, forming an oxide tail with voids left in the PdO tail as the Pd/PdO interface rapidly moves (marked by the white dashed line). When the PdO is reduced, the Pd/PdO interface migrates in the opposite direction, with no voids observed within the Pd phase. The cyclic oxidation and reduction processes further accelerate when the gas-flow rate reaches 0.28 sccm (**Figure 4e-4h**). This enhanced catalytic activity is likely due to improved heat convection at higher gas-flow rates, given the exothermic nature of methane combustion[57]. The Pd/PdO NPs become highly mobile and exhibit chaotic dynamic behaviour, differing from the interfacial bubbling situation. By tracking the evolution of PdO's projection area, we find that the redox timescale is approximately several seconds (**Figure 4i**). The system's heating power, temperature, and the $CO_2$ signal of MS exhibit synchronized periodic evolution (**Figure 4j**), indicating the formation of global oscillation whose period is much longer than the redox timescale of individual Pd NPs. Notably, at the micron- to meso-scale, all the Pd NPs appear to behave independently, resulting in an overall disordered domain. These observations suggest that there are no specifically ordered activities of single particles corresponding to the overall oscillatory dynamics, contrasting with the synchronization observed in foil catalysts[8,32].

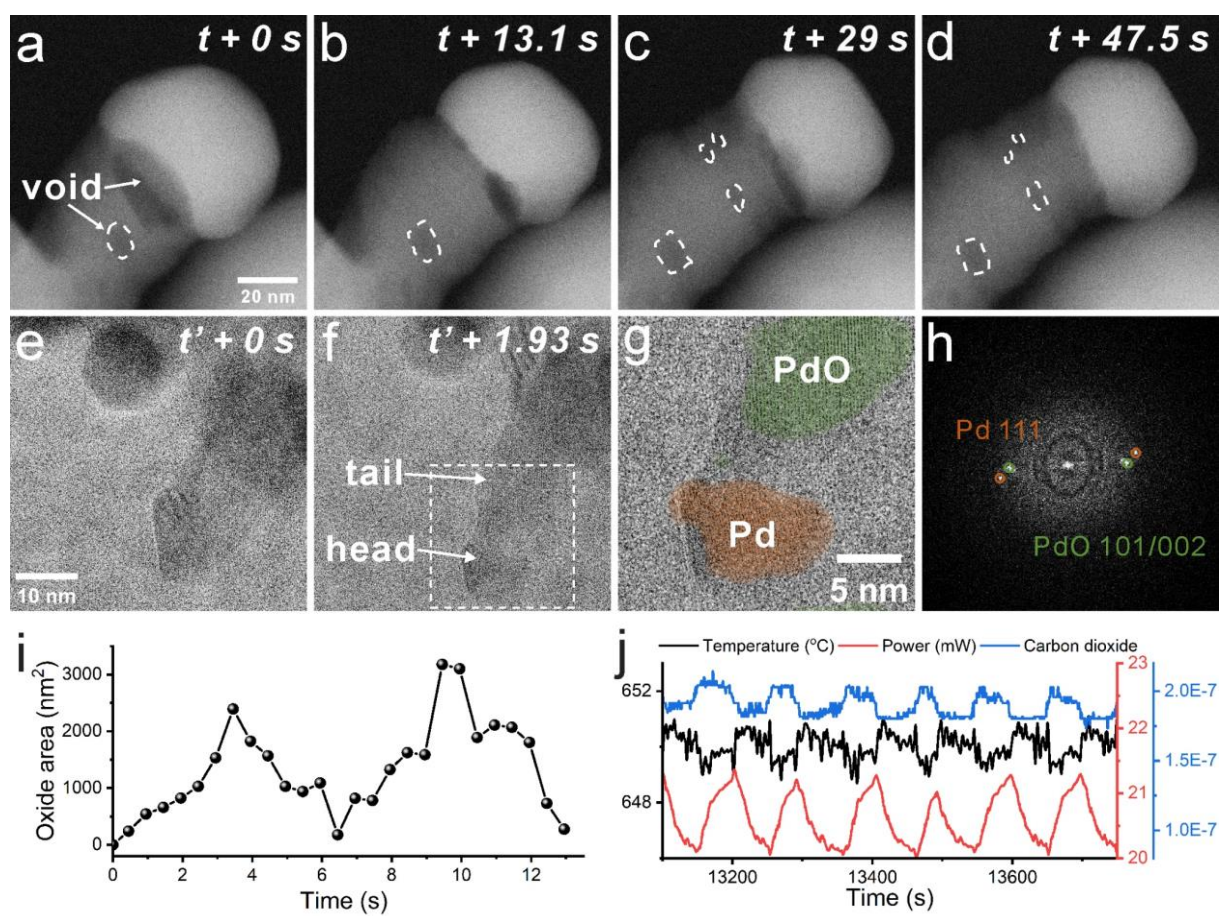

**Figure 4. Dynamics Variation of Pd Nanoparticles by Tuning Gas-Flow Rate.** (a-d) The interface migration of Pd oxidation at a $CH_4/O_2$ ratio of ~2/3 ($P_{total}$ = 902 mbar) and 650°C with a gas-flow rate of ~0.014 sccm. The white dashed arrow and lines highlight the blocked and interfacial voids during the oxidation process of Pd (300 kV, electron dose: 146.1 e$^-$/Å$^2$). (e-f) Oxidation of Pd nanoparticles during the redox process. The white arrows indicate the oxide tail and metal head (300 kV, electron dose rate: 1343.8 e$^-$/(Å$^2$·s)). (g) FFT-filtered image of the white rectangular area in (f) where the green and yellow lattice correspond to Pd and PdO,

respectively. (h) FFT of (g), Pd and PdO frequency dots are marked by the yellow and green circles respectively, which also represent the selected points for the mask filter of (g). (i) Statistical evolution of the oxide area of a single particle. (j) Macroscopic temperature, power, and $CO_2$ MS signal at a $CH_4/O_2$ ratio of ~2/3 ($P_{total}$ = 902 mbar) and 650°C with a gas-flow rate of ~0.28 sccm.

**Emergence of Collective Oscillatory Dynamics in Coupled Chaotic Systems**

To understand the timescale mismatch between the redox behaviour of individual NPs and the periodic oscillation of the global reaction, we recorded the evolution of lattice dynamics of Pd NPs through SAED patterns at the same 3μm-diametered region in a low magnification TEM image (**Figure 5a-5b**). The results show that the intensity of the Pd (111) diffraction ring oscillates and shares the approximate period with heating power and the $CO_2$ MS signal (**Figure 5c**). This indicates that global oscillation corresponds not to the uniform behaviour of individual NPs, but to the statistical distribution of Pd and PdO in the micron range. This scale gap suggests that thermal coupling between NPs exists locally and serves as a long-term correlation, adjusting the system's overall dynamics. This makes the emergence of dissipative structures (i.e. oscillations, spatial patterns, and waves) from chaotic agents[11,58,59].

The oscillatory dynamics emerging from multiple chaotic agents is known as weak synchronization[60]. This concept is supported by both theoretical frameworks and experimental observations in various biological systems[41,61,62], which exhibit characteristics analogous to NP groups used in heterogeneous catalysis. During $CH_4$ oxidation, Pd NPs show self-enhanced movement similar to the behaviour of living cells in a liquid environment[41]. The communication typically observed between cells is analogues to heat convection facilitated by the gas-flow. Higher flow rates enhance this interaction, fostering correlations between Pd NPs and enabling the formation of dissipative structures. Consequently, like a self-organized living system, the NPs are no longer isolated, and the system transitions into a periodic oscillatory state.

At high gas-flow rates, thermal interaction intensifies, and Pd NPs display more dynamic redox behaviour. The heat released during Pd oxidation to PdO in one particle is conveyed to its neighbouring area, increasing the temperature there. This temperature variation alters the chemical equilibrium of other Pd NPs[63], affecting the relative proportion of Pd and PdO. Conversely, the energy absorbed during PdO reduction is compensated by heat inflow, cooling the neighbouring area and promoting Pd oxidation.

Due to this thermal coupling and feedback mechanism, the overall redox behaviour in a local reactor domain exhibits oscillatory characteristics. The total contribution of this domain at the micron scale, as indicated by SAED patterns, can be viewed as an oscillator with well-defined periods and phases. Globally, the entire MEMS reactor comprises multiple oscillator-like domains. These oscillators collectively form a synchronized global pattern through thermal coupling across different mesoscale domains within the MEMS reactor. This process is aptly described by the Kuramoto model, a mean-field theory modified by coupling strength[64-66].

**Simulation of Global Oscillation**

With the above understanding, the weak synchronization approach to global periodic oscillation is further validated through spatial thermal field simulation. In this simulation, 400 heat sources are evenly distributed on a 100 × 100 grid plane to emulate Pd NPs. Each heat source is randomly assigned one of 36 different stochastically generated periodic heat release and absorption patterns, representing the fluctuating phase transition between Pd and PdO (**Figure 5d-5e**). The intensity of different lattice points reflects the temperature distribution in the mesoscale domain of the MEMS reactor, and their spatial average is regarded as the mean temperature.

To simulate thermal coupling, all sources in the grid interact with their neighbours. The interaction coefficient is modulated to reflect the gas-flow rate. When the interaction coefficient is high enough, the time-stacked evolution results demonstrate clear periodic oscillation of the entire plane (**Figure 5f**). Meanwhile, the dynamics of individual points remain chaotic (**Figure 5g**), with their timescales being shorter than the period of the global field oscillation (**Figure 5h**). Conversely, such an oscillation pattern does not exist when the interaction coefficient is very low.

The simulation aligns perfectly with experimental observations and highlights the importance of thermal coupling. In this context, the interaction coefficient represents the thermal coupling strength. When the coefficient is high, the thermal coupling is enhanced, leading to synchronized behaviour. Conversely, when the coupling is weak, the point heat sources remain isolated, preventing the formation of an overall ordered dynamic mode in the system.

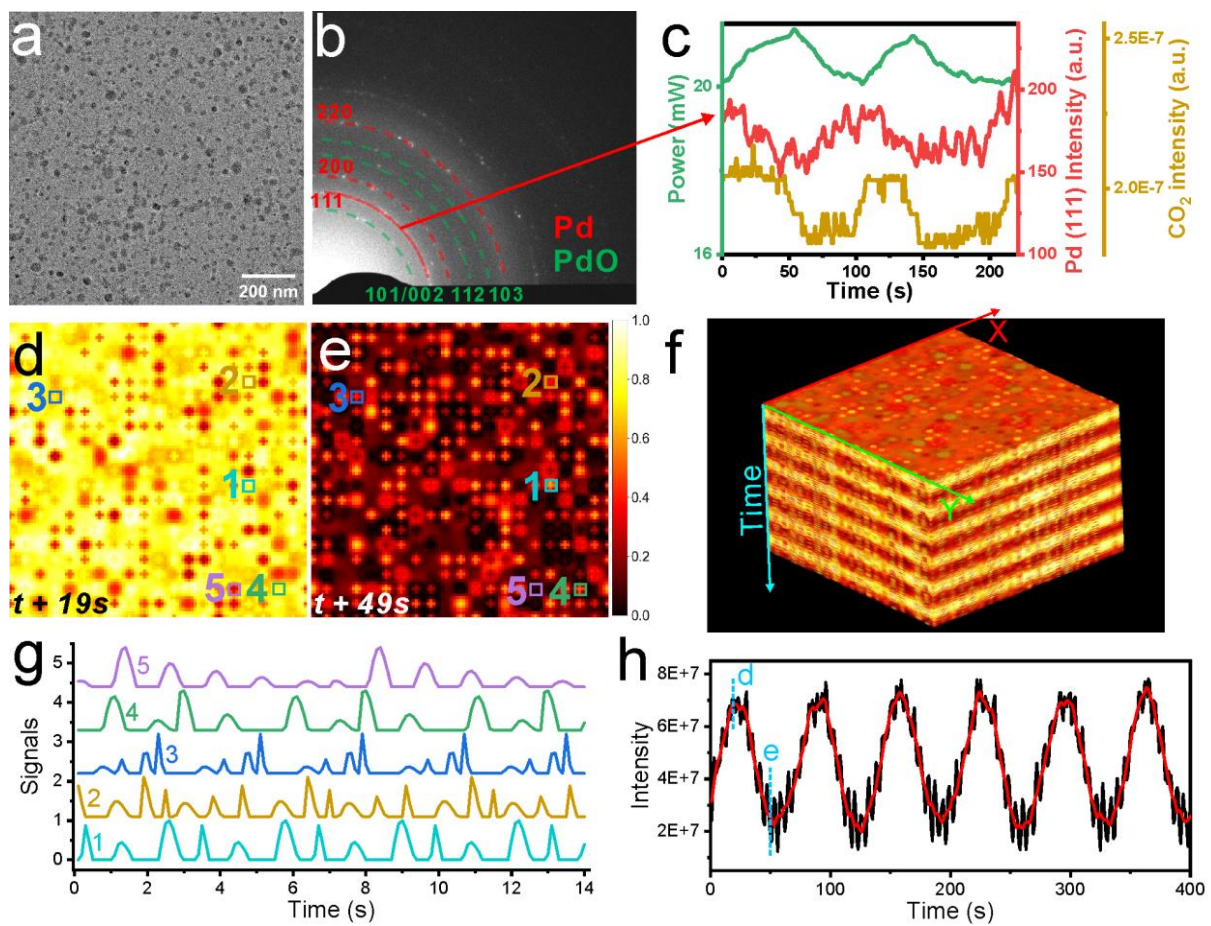

**Figure 5. Combining SAED with Systematic Macro-Oscillation and Thermal Field Simulation**. (a) TEM image showing the distribution of Pd NPs. (b) SAED pattern of the same area as (a). The peak of Pd (111) is marked by a red semi-transparent line, pointed to (c) by a red arrow and other different planes of Pd and PdO according to d-spacing, with red and green dashed lines, respectively. (c) Macroscopic power, Pd (111) intensity, and $CO_2$ MS signal showing similar periods. (a), (b) and (c) are at a $CH_4/O_2$ ratio of ~2/3 ($P_{total}$ = 902 mbar) and 650°C with a gas-flow rate of ~0.28 sccm (80 kV). (d-e) Simulation of the thermal field where 400 oscillatory sources are distributed in a 100 × 100 grid with a coupling constant value $\alpha$ = 0.2. The colour bar shows the normalized value of dots. (d) and (e) show two states of thermal field, as marked in (h) with azure colour. (f) The time stacking of thermal field simulation, showing the periodic oscillation over 400 s. (g) Time evolution of an individual point, with corresponding positions marked by coloured rectangles in (d-e). (h) Total intensity time evolution extracted from the whole region of (d-e), with the red curve smoothed using the Savitzky-Golay filter method with a 10-point window.

**Phase Transition from Apparent Steady State to Global Oscillations**

Beyond the significance of gas-flow rates, the crucial role of thermal coupling in the formation of global oscillation is verified by altering the reactor's temperature. By continuously increasing the temperature, we observed the evolution from an apparent steady state with minor changes in global signals to an oscillatory state (**Figure 6a-6b**). Between these two states, a transition state characterized by non-repeating oscillations, demonstrating chaotic characteristics, occurs within the temperature range from 550°C to 650°C. Above this range, the oscillations become more repeatable and stable. Conversely, below this range, the evolution curve shifts suddenly from steady to fluctuating states at a critical point. Since there is no additional latent heat effect apart from the emergence of chaotic oscillations, this transition exhibits features of a second-order phase transition[67-69]. Throughout the entire temperature increase, the system transitions from disordered to ordered phases, contradicting the general expectation of thermal fluctuations leading to increased disorder[70,71].

The observed temperature variation also aligns with our discussion based on the weak synchronization mechanism. According to the weak synchronization mechanism, the oscillation is more stable at higher temperatures because the reaction is expedited, accelerating heat production. Consequently, the thermal coupling between particles is enhanced, stabilizing the dissipative structures that arise from this coupling.

**Microkinetic Simulation of the Oscillation State**

To interpret the various dynamic phases with increasing temperature, we apply the reaction microkinetic simulations based on a reaction network of Pd-catalysed methane oxidation. By adjusting the reaction rate constant, both chaotic and stable waveforms observed experimentally can be replicated (**Figure 6c**). Furthermore, the kinetic curve in phase space, with variables of power and temperature of the microreactor, displays a limit cycle pattern (**Figure 6d-6e**), with blue arrows indicating the direction of time evolution, demonstrating the typical spatial and temporal order of dissipative structures[72,73].

This microkinetic calculation aligns with the prediction of the weak synchronization mechanism based on thermal coupling. Our explanation, supported by multi-scale *in situ* observations, clarifies the bottom-to-top mechanism of self-sustained oscillation through real-space coupling. Additionally, the oscillation waveform's sensitivity to environmental conditions and evolutionary trajectory highlights the chaotic and nonlinear characteristics of the reaction system. Therefore, the catalyst's reactivity can be modulated by the alternating oscillation mode of the system. Finally, a control experiment with an empty MEMS reactor shows no periodic evolution, ruling out the possibility that the systematic oscillation is caused by the reactor background.

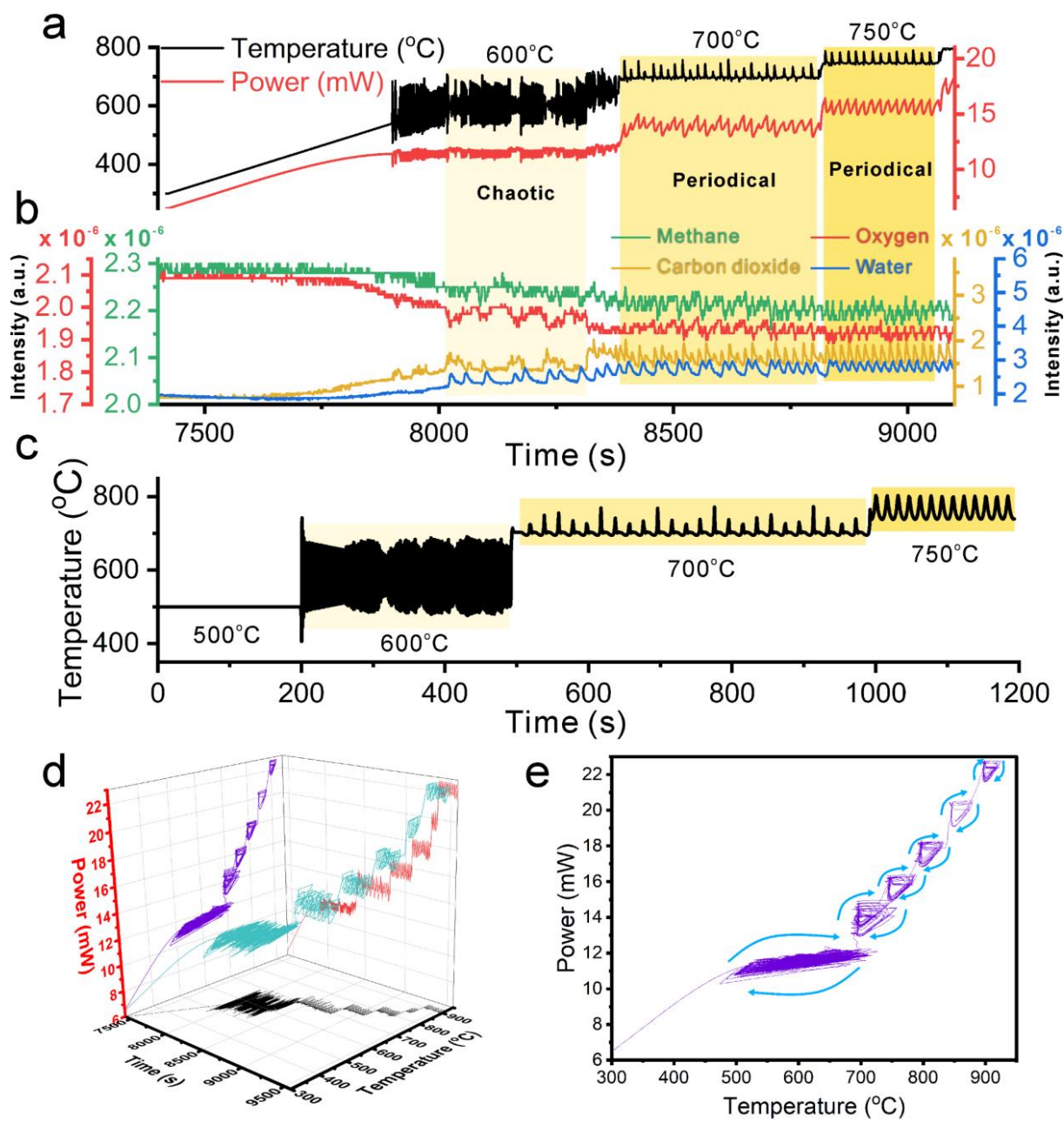

**Figure 6. Initialization of Global Oscillation and Corresponding Theory Simulation**. (a) Temperature and power of the MEMS chip, and (b) the MS results at a $CH_4/O_2$ ratio of ~1:1 ($P_{total}$ = 860 mbar) with a gas-flow rate of ~0.48 sccm during the continuous temperature variation experiment. All the mentioned parameters remain stable with minor perturbation until the temperature is approaching to 600°C. At 600°C, a chaotic oscillation pattern with stochastic periods arises. At 700°C, repeatable periodic oscillation occurs with minor fluctuations. Further

increasing the temperature to 750°C makes the oscillation period even more repeatable. (c) Temperature simulation at 600°C, 700°C, and 750°C under the same gas conditions as in (a). The yellow backgrounds in (a) and (c) indicate the temperature settings of 600°C, 700°C, and 750°C, respectively. (d) A 3D phase portrait in the same condition of (a) showing the relationship between time, temperature, and power, with projections in black, red, and purple, respectively. (e) A phase portrait showing power and temperature, with blue curved arrows indicating the direction of time evolution.

**Conclusion and Outlook**

In this work, we report on the self-sustained oscillations of methane combustion catalysed by Pd NPs using complementary *operando* methodologies. We demonstrate that gas-flow-induced thermal coupling transitions of the system from an apparent steady state to a global oscillatory state. At low gas-flow rates, water accumulation at the Pd/PdO interface leads to repeated crystal facet reconstruction at the individual particle level, with no macroscopic effects due to inefficient heat transfer. Higher gas-flow rates accelerate interfacial reactions, causing the Pd/PdO interface to migrate and exhibit self-enhanced movement, and resulting in global parameter oscillations.

Operando observations reveal that global oscillation is not driven by individual particle dynamics but by weak synchronization through spatial coupling. This coupling transforms local disorder into periodic collective dynamics in micron-scale domains, producing global oscillation. Temperature variation experiments show that the transition from a steady state to an oscillatory state involves intermediate chaotic oscillation, essential for modulating oscillation behaviour.

Our work combines multiple experimental technologies to achieve temporally resolved observations of catalyst behaviours across different spatial scales, which is rarely done in previous research[26,35,74-77]. The results complement each other, forming a bottom-up picture of the emergence of ordered oscillation patterns. Objects at different spatial scales exhibit varying degrees of order and dynamic patterns, consistent with Anderson's philosophy on emergence and complexity in dissipative systems[78]. This complexity in the Pd NPs system, arises from thermal effects, numerous degrees of freedom, and variations in gas-flow coupling strength.

Following this philosophy and utilizing the complementary *in situ/operando* methods, we investigate the principles governing the emergence of complex patterns, potentially leading to the creation and modulation of diverse dissipative structures in physical and chemical systems under nonequilibrium states. Oscillation, a representative type of Prigogine's dissipative structures, has been observed in active matter like living cells[17,18], molecular motors[42,43], and soft matter machines[79,80]. These systems consume local free energy and display self-organized patterns[17,18,81], analogous to our heterogeneous catalyst systems. The existence of dissipative structures allows us to create order from disorder by changing environmental variables, as shown in our work. Our findings connect nanoparticle activity with the evolution of global variables, emphasizing the critical role of coupling strength modulated by environmental parameters. This framework is likely applicable to other catalysts and reactions. Practically, understanding oscillation properties can improve productivity and efficiency in the chemical industry by adjusting reaction modes. Additionally, the refreshing effect of oscillation cycles can aid in developing highly efficient never-deactivating catalysts[82,83].

**Acknowledgements**

This work was mainly supported by the National Natural Science Foundation of China under grant no. 12027804. We acknowledge the Center for High-Resolution Electron Microscopy of ShanghaiTech University for use of the electron microscope. J.H. and W.L thank the National Natural Science Foundation of China (22221003), the Strategic Priority Research Program of the Chinese Academy of Sciences (XDB0450102), Innovation Program for Quantum Science and Technology (2021ZD0303302), the Fundamental Research Funds for the Central Universities (20720220009). The authors also gratefully thank Supercomputing Center of University of Science and Technology of China. The authors thank BL02B at the Shanghai Synchrotron Radiation Facility supported by National Natural Science Foundation of China under contract no. 11227902.